\documentstyle[preprint,aps]{revtex}
\draft

\newcommand{\ba}{\begin{eqnarray}}
\newcommand{\ea}{\end{eqnarray}}

\begin{document}

\title{Magnetic Dipole Sum Rules for Odd-Mass Nuclei}
\author{ J.N. Ginocchio$^{1,3}$  and A. Leviatan$^{2,1,3}$}
\address{$^{1}$~Theoretical Division, Los Alamos National Laboratory, Los
Alamos, New Mexico 87545, USA}
\address{$^{2}$~Racah Institute of Physics, The Hebrew University,
Jerusalem 91904, Israel}
\address{$^{3}$~European Centre for Theoretical Studies in Nuclear Physics
and Related Areas (ECT*), I-38050 Villazano, Trento,
Italy}
%\date{\today}

\maketitle

\begin{abstract}
Sum rules for the total- and scissors-mode M1 strength in 
odd-A nuclei are derived within the single-j interacting boson-fermion
model. We discuss the physical content and geometric interpretation
of these sum rules and apply them to $^{167}$Er and $^{161}$Dy. We 
find consistency with the former measurements but not with the latter.
\end{abstract}


The orbital magnetic dipole scissors mode \cite{boh84} has by now been 
established experimentally as a general phenomenon in deformed 
even-even nuclei \cite{ric95kne96}. The M1 strength systematics 
and deformation dependence have been extensively measured
and corroborated in a variety of sum rules \cite{gin91,sum}.
Within the interacting boson model (IBM) \cite{iac87}, 
a sum rule \cite{gin91} has related this strength  
to the number of quadrupole bosons in the ground
state of the even-even target nucleus, $\langle 0|N_d|0 \rangle$,
\ba
\sum_f B(M1:0^+ \rightarrow 1^+, f) = \frac{3}{4 \pi} \, g_v^2 \,
\frac{6N_{\pi}N_{\nu}}{N(N-1)} \; \langle 0|N_d|0 \rangle \; ~.
\label{eq_oldsr}
\ea
Here $g_v = (g_{\pi} - g_{\nu})$ , $g_{\rho}$ are the proton 
$(\rho = \pi)$ and neutron $(\rho = \nu)$ boson g-factors and
$N_{\rho}$ the corresponding boson numbers, $N=N_{\pi}+N_{\nu}$.
For deformed nuclei $\langle 0|N_d|0 \rangle $
can be expressed in terms of the deformation 
determined from B(E2) values, and 
the measured M1 strength was shown
to be in good agreement with this
sum rule \cite {vnc}. 
A survey \cite{isa89fra91} of scissors states in deformed odd-mass nuclei 
within the framework of the interacting boson-fermion model (IBFM) \cite{ibfm}
predicted strong fragmentation and sizeable symmetric (single-particle
dominated) and non-symmetric (scissors) M1 strength.
Initial measurements on odd nuclei (Dy, Gd, Tb) 
indicated
missing B(M1)$\uparrow$ 
strength compared to the strength observed in the neighboring even-even 
nuclei \cite{kn1,kn2,kn3}. On the other hand, in $^{167}$Er 
these M1 strengths were found to be comparable, with appreciable 
contribution coming from higher energies
not accessed in previous experiments \cite{Er167}.
Inspired by these puzzling results we have derived 
sum rules for the total- and scissors mode M1 strength
in odd nuclei within the IBFM.
We focus on odd nuclei with the fermion
in an orbital with single-nucleon angular momentum j, which, 
for practical purposes, means that our sum rules are applicable to
nuclei for which the ground state
has the fermion filling the intruder (abnormal) orbit.
In general, we find that the sum rules measure two 
quantities, the average value of $N_d$ in the ground state and the 
average relative orientation of the single-particle and total angular 
momenta. We apply the formalism to $^{161}$Dy and $^{167}$Er having an odd
neutron in the $1i_{13/2}$ orbital.

We begin by defining the weak coupling basis in which a nucleon 
with angular momentum $j$ is 
coupled to a bosonic wave-function 
$\ |\ [\alpha,F, F_0, L],j; J, M \rangle $ specified by the $F$-spin 
\cite {ots78} $F$, its projection, $F_0 = (N_{\pi} - N_{\nu})/2$, 
the angular momentum of the IBM core $L$, and additional labels $\alpha$.
$J$ is the total angular momentum (projection $M$)
resulting from the coupling of $L$ and $j$.
The initial ground state wave function will then
be a linear combination of these states, 
\ba
 | i; J \rangle =
\sum_{L} C_{i,L}
|\ [\alpha_{i},F_m, F_0, L],j; J M\rangle
\label{weakwf}
\ea
with $F_{m}=N/2$ the maximal
$F$-spin and the label $i$ indicates all
quantum numbers that may be needed to specify uniquely 
the initial state. Throughout this discussion we assume that the
boson $F$-spin is conserved. 
The magnetic dipole operator is given by,
\ba
T_m = g_{\pi} \hat L_{\pi,m} + g_{\nu} \hat L_{\nu,m} + g_{j} \hat
j_{m} =
g \hat J_{m} + g_{v} {\hat L}_{v,m} + g_{F} \hat
j_{m}\;\;,
\label{eq_mag}
\ea
where $\hat J_{m} = \hat L_{m} +  \hat j_{m}$,
$\hat L_{m} = \hat L_{\pi,m} + \hat L_{\nu,m}$ and $\hat j_{m}$ are the
total-, IBM core-, and single-nucleon angular momentum operators 
respectively, ${\hat L}_{v,m} = ( \hat L_{\pi,m} - \hat L_{\nu,m} )/2$
with $\hat L_{\rho,m}$ the individual boson angular momenta operators, 
$g_{j}$ is the nucleon g-factor, and
$g = (g_{\pi} + g_{\nu})/2$, $g_{v} = g_{\pi} - g_{\nu}$,  
$g_{F} = g_{j} - g$. The operators $\hat J_{m}$, $\hat L_{m}$ and 
$\hat j_{m}$ are F-spin scalars ($F=F_0=0$) and contribute only to 
symmetric $\rightarrow$ symmetric transitions $(F_m\rightarrow F_{m})$.
The $\hat L_v$ operator is F-spin vector 
($F=1,F_0=0$) and contribute also to symmetric $\rightarrow$ non-symmetric 
(scissors) transitions $(F_m\rightarrow F_m-1)$ \cite{isa89fra91}.
The total B(M1) strength from the ground 
state to all final states is given by,
\ba
\sum_f B(M1:i \rightarrow f) = \frac{3}{4 \pi}  
\frac{\sum_f |\langle i; J||T||f; J_f \rangle|^2 }{(2J + 1)} 
= \frac{3}{4 \pi} \langle i; J|\ T\cdot T\ |i; J \rangle\;\;,\;
\label{eq_bm1}
\ea
where the dot in $T\cdot T$ denotes a scalar product and 
$\langle i; J||T||f; J_f \rangle$ is a reduced matrix element (r.m.e.).
By using tensor operator identities \cite{russ} and the Wigner-Eckart
theorem in F-spin space, we can evaluate the matrix elements of 
$T\cdot T$ in the weak coupling basis. 
Specifically, $T\cdot T$ is a sum of terms of the form $\hat L\cdot\hat L$,
$\hat j\cdot\hat j$, $\hat L\cdot\hat j$, $\hat L\cdot \hat L_{v}$,
$\hat L_{v}\cdot \hat j$ and $\hat L_{v}\cdot\hat L_{v}$. States in 
the weak coupling basis are eigenstates of the first three terms. 
The matrix elements of $\hat L_{v}\cdot \hat j$ reduce to a 
r.m.e. of the angular momentum $\hat j$ in the fermion space (which is known) 
times the r.m.e. of $\hat L_{v}$ between boson states with 
$(F_{m},F_0)$. The bosonic matrix element is related by the Wigner Eckart 
theorem to a matrix element between states with $(F_m,F_0=F_m)$ for
which $\hat L_{v}$ become the total boson angular momentum in 
the boson space, whose matrix elements are diagonal and known. 
Similar F-spin reduction appear in the matrix elements of 
the F-spin vector $\hat L\cdot \hat L_{v}$ term.
Likewise, we can evaluate 
the ${\hat L}_v \cdot {\hat L}_v$ contribution by decomposing it to $F$-spin
components ($F=2,F_0=0$) and ($F=F_0=0$), in the same way that was done for
even-even nuclei in \cite{gin91}. 
The relevant $F$-spin Clebsch Gordan Coefficients
(C.G.C.) induce particular $N_{\rho}$-dependence for each term.
The resulting matrix elements of $T\cdot
T$ in the weak-coupling basis are diagonal in L and
by averaging them over $\sum_{L}C_{i,L}^2$ we can evaluate 
the right hand side of Eq. (\ref{eq_bm1}).

	Consider first the $M1$ strength from the ground to the 
scissors states. This involves symmetric$\rightarrow$non-symmetric transitions
with a change of $F$-spin: $F_{m}\rightarrow F_{m}-1$, and hence are 
induced only by the isovector operator $g_v\hat L_v$. 
Consequently, their contribution to the summed
strength is proportional to the $F$-spin C.G.C. 
$g_{v}^2(F_{m}-1,F_0;1,0|F_{m},F_0)^2 = 
2g_{v}^2N_{\pi}N_{\nu}/N(N-1)$. 
This dependence provides the signature needed to identify the contribution
of these transitions to the right hand side of Eq. (\ref{eq_bm1}), 
and thus leads to the following sum-rule for the M1 strength from the ground 
state to the scissors mode ($sc$) in odd-A nuclei,
\ba
\sum_{f} B(M1: i \rightarrow sc, f) =
\sum_{f} B(M1: i \rightarrow f)_{core} \; ~.
\label{sc_sum}
\ea
Here 
\ba
\sum_{f} B(M1:i \rightarrow f)_{core} & = & 
\frac{3}{4 \pi} \, g_v^2 \, \frac{N_{\pi}N_{\nu}}{N(N-1)} 
\big [\, 6 \langle \,N_d\,\rangle -
\frac{\langle \,L(L+1) \,\rangle}{N} \, \big ] ~,
\label{eq_core}
\ea
and
\begin{eqnarray}
\langle \, N_d \, \rangle & = & \sum_{L} C_{i,L}^2
\langle \,  [\alpha_i,F_m, F_0, L],j;JM \ |\,  N_d \, 
|\,  [\alpha_i,F, F_0, L], j; JM \, \rangle ~,
\nonumber\\
\langle \, L(L + 1) \, \rangle &=& \sum_{L} 
C_{i,L}^2 \ L ( L + 1).
\label{eq_ndl}                              
\end{eqnarray}
The first term in expression (\ref {eq_core}) is similar to that
in Eq. (\ref{eq_oldsr}), except that in the latter
$\langle 0|N_d|0 \rangle$ is the
average number of quadrupole bosons in the ground state of the even-even 
nucleus with boson angular momentum zero ($L=0$), 
whereas $\langle N_d\rangle$ in 
Eqs. (\ref {eq_core})-(\ref{eq_ndl}) is the
average number of quadrupole bosons in the core of the neighboring 
odd nucleus which will have an admixture of boson angular
momenta. However, for deformed nuclei in the large N limit, 
$\langle N_d\rangle/N$ is
independent of the boson angular
momentum to order ${\cal O}(1/N)$\cite {kuy95}. The $L$-dependent
correction to $\langle N_d\rangle/N$ of ${\cal O}(1/N^2)$ has the same 
dependence on the average angular momentum square, $\langle L(L+1)\rangle$, 
as the second term in (\ref {eq_core}), and reduces the magnitude of this 
${\cal O}(1/N^2)$ correction. Hence we suggest
that the two terms in Eq. (\ref{eq_core}) constitute the contribution of the 
core to the total B(M1) and can be calculated in the way outlined in
\cite{vnc} to order ${\cal O}(1/N)$ using the deformation of the odd nucleus.
To this order, the sum rule in Eq. (\ref{sc_sum}) states that
the summed M1 strength from the ground to the scissors mode in an odd
nucleus exhibits quadratic dependence on the Bohr-Mottelson quadrupole
deformation \cite{vnc} and is equal to the summed 
$B(M1:0^{+}\rightarrow 1^{+})$ 
strength in the neighbouring even-even nucleus. 
Adapting the criteria of \cite{piet95} for assigning measured M1 strength
to the scissors mode, we obtain the values for $B(M1)_{core}$ which are  
$2.42(0.18)$ $\mu_N^2$ from $^{160}$Dy \cite{Dy160} 
and $2.67(0.19)$ $\mu_N^2$ from $^{166}$Er \cite{Er166}. 
The measured scissors strength is $0.71(0.10)$ $\mu_N^2$ 
in $^{161}$Dy \cite{kn2} and $3.14(1.12)$ $\mu_N^2$ in 
$^{167}$Er \cite{Er167} assuming M1 transitions. Referring to
the sum rule of Eq. (\ref{sc_sum}), these results
indicate significant lack of M1 scissors strength in $^{161}$Dy and 
consistency, within the experimental errors, for $^{167}$Er. 

To obtain a sum rule for the total M1 strength in odd nuclei we
need to consider the contributions of all terms in Eq. (\ref{eq_bm1}).
We note, however, that unlike even-even nuclei, there is a magnetic
dipole transition to the ground state proportional to the magnetic moment.
This elastic transition is not measured in 
$(e,e')$ and $(\gamma,\gamma')$ experiments, which employ continuous wave
beams (Bremsstrahlung); hence we subtract it and obtain
\ba
\sum_{f\ne i} 
B(M1:i \rightarrow f) = \frac{3}{4 \pi}\ \left [ \langle i; J|\
T\cdot T\ |i; J \rangle\ - \frac{|\langle i; J|| T ||i; J
\rangle|^2}{(2J+1)}\ \right ]\ .\;\;
\label{eq_bmf}
\ea
The subtraction in Eq. (\ref{eq_bmf}) results in partial cancellation
of terms and we finally arrive at the following sum rule
\ba
\sum_{f\ne i} B(M1:i \rightarrow f) = \sum_{f} B(M1:i \rightarrow
f)_{core} +\frac{3}{4 \pi}\ \big (g_{j}-\bar g\big)^2
j(j+1)\Bigl [1 -\langle\, \cos\theta\, \rangle^2 \Bigr ] \, ~,
\label{eqF}
\ea
where $\bar g = (N_{\pi}g_{\pi} + N_{\nu}g_{\nu})/N$ is the 
weighted boson g-factor and
\ba
\langle\, \cos\theta\, \rangle = 
\frac {j(j+1) +J(J+1) - \langle \,L(L+1)
\,\rangle}{2\sqrt{j(j+1)J(J+1)}} \, ~.
\label{eq_cos}
\ea
The right hand side of Eq. (\ref{eq_cos}) is the semi-classical expression 
\cite{russ} for the cosine of the angle $\theta$ between the vectors
$\vec{j}$ and $\vec{J}$ (see Fig.~1). The sum rule, therefore,
provides information on the relative orientation of these angular
momenta which, in turn, as seen from Eq.~(\ref{eq_cos}), depends on the 
average angular momenta square $\langle\,L(L+1)\,\rangle$ of the core.
From (\ref{eqF}), we see that the single-nucleon contribution
to the B(M1) vanishes for both total alignment and
anti- alignment ($\theta = 0, \pi$), and is maximum when $\vec{j}$ and
$\vec{J}$ are perpendicular ($\theta =\pi/2$). The particular
$j(j+1)\Bigl [1 -\langle \,\cos\theta\, \rangle^2 \Bigr ]$ dependence
is intuitively understood \cite{don87} from the geometry of
the angular momenta shown in Fig.~1. For the magnetic transition
strength only the component of $j$ perpendicular to $J$ 
($j_{\perp}=j\sin\theta$ ) is effective (oscillating dipole as $\vec{j}$ 
precesses about $\vec{J}$ ), whereas the parallel component
($j_{\parallel}=j\cos\theta$) contributes only to the static moment
(see Eq. (\ref{eq_mom}) below).
The core contribution in Eq. (\ref{eqF}) is determined by the same procedure
discussed for the scissors mode sum rule of Eq. (\ref{sc_sum}).
The parameters $\bar{g}$
and $\langle\, \cos\theta \,\rangle$ can be determined from
the magnetic moments of the odd nucleus ($\mu_J$) and
the neighbouring even-even nucleus ($\mu_L$),
\begin{eqnarray}
\mu_L &=& \bar{g}L ~,
\nonumber\\
 \mu_J &=& \Bigl [ \,\bar g + (g_{j} - 
\bar g)\sqrt{\frac{j(j+1)}{J(J+1)}}\
\langle \,  \cos\theta \, \rangle \,  \Bigr ] \ J ~.
\label{eq_mom}
\end{eqnarray}
We take $\ g_j$ to be the Schmidt value
$g_{j,Schmidt}$ or the quenched 
Schmidt $ 0.7 g_{j,Schmidt}$ \cite{isa89fra91}.

For weak coupling, the ground state has $L = 0$ and $J = j$, so that 
$ \langle \, \cos\theta \, \rangle = 1$ and hence both the total- and scissors
M1 strengths are equal to $B(M1)_{core}$.
However, in general the sum rule in Eq. (\ref {eqF}) is an upper limit on the 
total B(M1) strength because the basis states of bosons coupled to a 
single-nucleon will be over-complete since the
bosons represent coherent pairs of fermions some of which are occupying the
single-nucleon orbital j.
Hence the sum rule which includes only Pauli allowed states (PA) will be in
reality given by,
\ba
\sum_{f\ne i} B(M1:i \rightarrow f)_{PA} = \sum_{f\ne i} B(M1:i \rightarrow
f) - \sum_{f\ne i} B(M1:i
\rightarrow f)_{PF}
\label{eq_pa}
\ea
To calculate the Pauli forbidden (PF) strength we need a model
which we presently take to be the strong coupling limit of the 
IBFM \cite {ibfm,ami}. In this limit, the amplitudes in 
Eq. (\ref{weakwf}) are proportional to a C.G.C. 
$C_{K,L,j,J} = \sqrt{2(2L+1)/(2J+1)}\,(L,0;j,K|J,K)$ with $L$ even 
and $K$ is the projection of $j$ (and of $J$) along the symmetry 
axis (the corresponding projection of $L$ is zero).
We can then calculate
the required averages by using the fact that 
the expression for $\cos\theta$ in Eq. (\ref{eq_cos})
can be related to a 6-j symbol,
\ba
\langle \, \cos\theta \, \rangle = 
(-1)^{j+J+L+1}\sqrt{(2j+1)(2J+1)}\  \, \langle
\,
\Bigl \{
\begin{array}{ccc}
L & j & L \\
1 & J & j
\end{array}
\Bigr \}
\, \rangle
\label{eq_sixj}
\ea
and then use well-known identity for summing a 6-j times two C.G.C.
\cite {russ}. We find,
\ba
\langle \, \cos\theta \, \rangle = \frac{2K^2 + (-1)^{j-J} (J+1/2)(j+1/2)
\delta_{K,1/2}}{\sqrt{J(J+1)j(j+1)}}.
\label{eq_sc}
\ea
The transitions $K\rightarrow (K -1)$ are Pauli forbidden which 
gives ($K\ne 1/2$),
\ba
\sum_{f\ne i} B(M1:i,K \rightarrow f,K-1)_{PF} = \frac{3}{8\pi} 
(g_{j} - \bar g)^2 \,\Bigl [\,j(j+1) - K (K-1 )\,\Bigr ] \; ~.
\label{eq_pf}
\ea

We have evaluated the Pauli-corrected total M1 sum rule for the nuclei 
$^{161}$Dy ($j = 13/2$, $J = 5/2$, $K = 5/2$) and
$^{167}$Er ($j = 13/2$, $J = 7/2$, $K = 7/2$), taking the experimental
$B(M1)_{core}$ values as before from the neighboring even-even nuclei 
and using the total strength identified (assuming M1 transitions)
\cite{kn2,Er167}. From the magnetic moments 
of the first $2^+$ states in $^{160}$Dy and $^{166}$Er
we determine $\bar g  = 0.362\,\mu_N$ and $0.318\, \mu_N$, 
respectively. We then determined 
$\langle \, \cos\theta \,\rangle$ from the magnetic moments 
of $^{161}$Dy ($\mu_J = -0.480\, \mu_N$) 
and $^{167}$Er ($\mu_J = -0.56385\, \mu_N$)
using the bare Schmidt value of $g_{j,Schmidt} = -0.2943\,\mu_N$ 
for $1i_{13/2}$ neutron orbital, and the quenched value, 
$0.7 g_{j,Schmidt} = -0.2060\,\mu_N$. 
The results are summarized in Table I. We see that, whereas the
$^{167}$Er sum rule 
is consistent, within the
experimental error, with the measured value of the total M1 strength, for
$^{161}$Dy there is a large amount of missing strength, suggesting that
$^{161}$Dy has ``unexpected properties'', not
$^{167}$Er as implied by the title of \cite {Er167}.
The missing scissors M1 strength in $^{161}$Dy may reside at higher
energies (as encountered in $^{167}$Er \cite{Er167}) while  
the missing total M1 strength (single-particle dominated) may also be 
at lower energies where the increased background limits the 
experimental sensitivity \cite{kn3}. There are initial indications
that a fluctuation analysis of the spectra can be used to fix the
unresolved strength in the background and leads to comparable strengths
\cite{cosel}.

The sum rules reported in this work rely on good $F$-spin symmetry. 
The validity of this assumption and the implications
of breaking this symmetry on M1 transitions in odd nuclei
was elaborated in \cite{isa89fra91}, and shown that in the rare-earth region
the effect is small. The utility of the large-N approximation was
throughly investigated by the $1/N$ technique (see \cite{kuy95} and
references therein).
Although we have confined the discussion to a single-j model space 
for the odd fermion (hence specific odd-mass nuclei), similar formulas 
should hold for a situation of pseudospin symmetry.
Multi-j formulation
of similar M1 sum rules requires confronting 
more allowed terms (and parameters) in the M1 operator and several 
$j$-amplitudes in the fermion wave-function.

In summary, we have derived sum rules for magnetic dipole
transitions in odd nuclei with a nucleon in a single
spherical orbit, $j$, assuming F-spin symmetry and large-N approximation
for the boson core.
The summed M1 strength from the ground to the scissors mode in such nuclei 
depends on the average number of quadrupole bosons in the ground state. 
This quantity, for large $N$, exhibits quadratic 
dependence on the Bohr-Mottelson quadrupole deformation and can be
determined from the scissors M1 strength in the neighboring even-even 
nucleus or from the deformation of the odd nucleus to ${\cal O}(1/N)$.
The total M1 strength depends in addition also on the average 
of the cosine of the angle between the odd nucleon angular momentum and
the total angular momentum. This quantity is related to the average of the
core angular momentum, $\langle\,L(L+1)\,\rangle$, and can be determined 
from the magnetic moments of the target and the
neighboring even-even nucleus. In general this sum rule will be an
upper limit from which the strength to Pauli forbidden states needs to be 
subtracted. In well-deformed nuclei the strong coupling limit can be 
used for that purpose. In comparing with the known B(M1) strength, we find
consistency in $^{167}$Er for the scissors- and for the total strength.
However, we conclude that a
substantial portion of the magnetic dipole scissors and total strengths 
are missing in $^{161}$Dy. 
Clearly, more measurements both at higher and lower energies are needed
in $^{161}$Dy to see if the predictions of the sum rules are satisfied.
More and detailed calculations are needed to understand M1 properties
in odd-mass nuclei. In particular,
further theoretical attention is needed for 
including Pauli corrections in estimates of M1 strength and for understanding 
the different fragmentation patterns observed in different odd nuclei.

We are indebted to U.~Kneissl, P.~von Neumann-Cosel and F.~Iachello 
for helpful discussions. We thank the European Centre for Theoretical 
Studies in Nuclear Physics and Related Areas (ECT*) for its kind 
hospitality during a stay from which this article originated.
This work was supported in part by a grant from the Israel Science Foundation
and by the U.S. Department of Energy.

\pagebreak
\begin{table}
\centering
\caption[]{\small
The total B(M1) strength in units of $(\mu_{N}^2)$ for 
$^{161}$Dy and $^{167}$Er for bare and quenched Schmidt $g_j$ values 
are tabulated. 
$\langle\,\cos\theta\,\rangle$ is defined
in (\ref{eq_cos}),  
$\sum B(M1)$ is the total contribution of the sum rule (\ref{eqF}), 
$\sum B(M1)_{PF}$ is the contribution
of the Pauli forbidden states in the strong coupling limit (\ref{eq_pf}) ,
$\sum B(M1)_{PA}$ is the total strength of the Pauli
allowed states (\ref{eq_pa}), and $\sum B(M1)_{exp}$ are the measured
values given in \cite{kn2,Er167}, respectively. Choice of parameters
is discussed in the text.
\normalsize}
\vspace{18pt}
\begin{tabular}{ccccccc}
Nucleus & $g_j $ & $\langle\, \cos\theta\, \rangle$ &  $\sum B(M1) $ &
$\sum B(M1)_{PF}$ &
$\sum B(M1)_{PA} $ &
$\sum B(M1)_{exp} $ \\
$^{161}$Dy & $g_{j,Schmidt}$ & \ 0.358 & \ 6.78 (0.18) &
\ 2.31 & \ 4.47 (0.18) & \ 0.88 (0.13)\\
$^{161}$Dy & 0.7$g_{j,Schmidt}$ & \ 0.413 & \ 5.53 (0.18) & \ 1.73 & \ 3.80
(0.18) & \  0.88 (0.13) \\
$^{167}$Er & $g_{j,Schmidt}$ & \ 0.445 & \ 6.16 (0.19) & \ 1.78 & \ 4.38 (0.19)
& \ 3.49 (1.15)\\
$^{167}$Er & 0.7$g_{j,Schmidt}$ & \ 0.520 & \ 5.00 (0.19) & \ 1.31 & \  3.69
(0.19)  & \  3.49 (1.15)\\
\end{tabular}
\end{table}

\clearpage
\begin{figure}
\caption{Angular momenta for a particle coupled to an axially-symmetric 
core. $J$, $L$ and $j$, are the total-, core-, and single-particle angular 
momenta respectively. $K$ is the projection of $j$ (and of $J$) on the symmetry
axis.}
\end{figure}

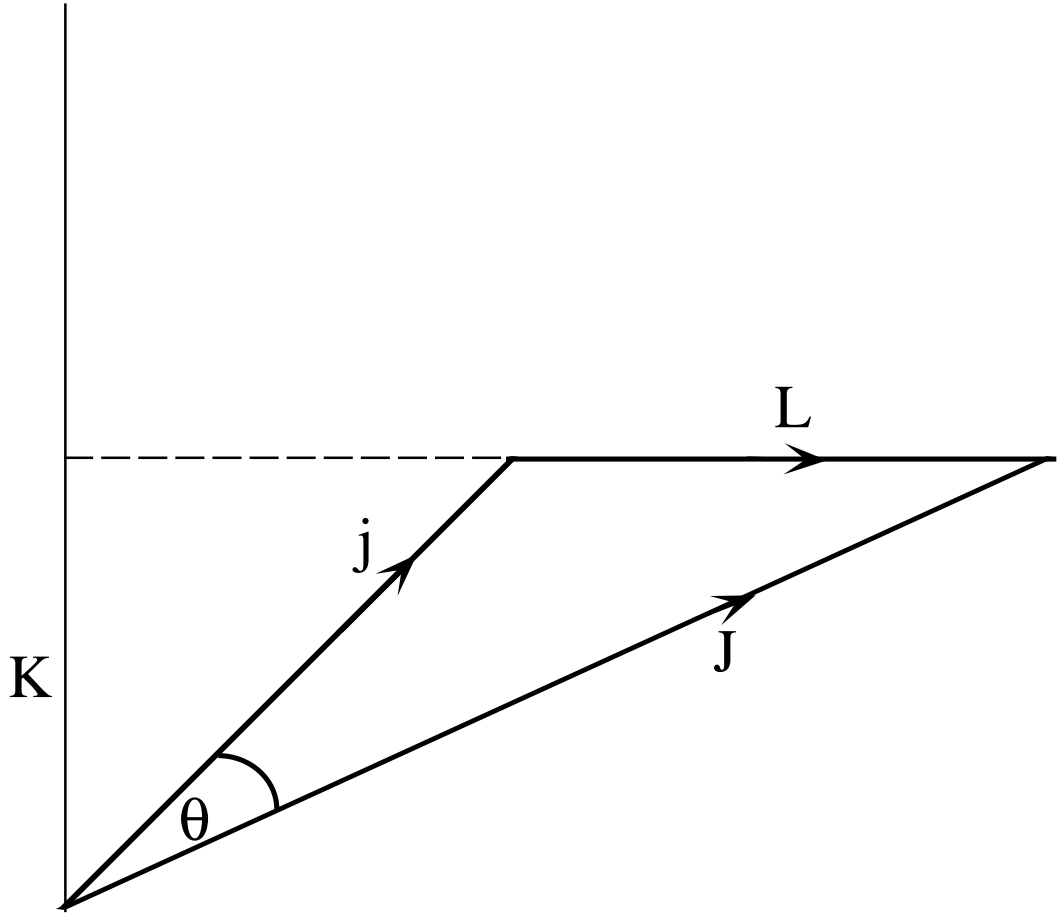
\end{document}